\documentclass[3p,twocolumn,authoryear]{elsarticle}
\pdfoutput=1  

\usepackage[T1]{fontenc}
\usepackage[utf8]{inputenc}
\usepackage[english]{babel}

\usepackage[lined,ruled,oldcommands]{algorithm2e} 

\usepackage[pdftex]{color}
\usepackage[pdftex,colorlinks=true]{hyperref}

\usepackage{amsfonts}
\usepackage{amsmath}
\usepackage{amssymb}
\usepackage{amsthm}

\usepackage{subfig}

\newcommand{\Z}{\mathbb{Z}}
\newcommand{\med}{\mathop{\mathrm{med}}}

\usepackage{ifdraft}

\journal{\ifdraft{DRAFT, NOT SUBMITTED}{Journal of Theoretical Biology}}

\newcommand{\Algoref}[1]{\hyperref[#1]{Algorithm~\ref*{#1}}}
\newcommand{\Figuref}[1]{\hyperref[#1]{Figure~\ref*{#1}}}

\begin{document}

\begin{frontmatter}

\title{Coupling methods for efficient simulation of spatial population dynamics}

\author{Ilmari Karonen\corref{ikaronen}}
\ead{ilmari.karonen@helsinki.fi}
\address{Department of Mathematics and Statistics, University of Helsinki, Finland}

\cortext[ikaronen]{\textit{Address for correspondence:} Department of
  Mathematics and Statistics, University of Helsinki, PO Box 68, FI-00014,
  Finland. Phone: +358-41-456 3263, fax: +358-9-1951 1400.}

\begin{abstract}
  Coupling is a widely used technique in the theoretical study of interacting
  stochastic processes.  In this paper I present an example demonstrating its
  usefulness also in the efficient computer simulation of such processes.  I
  first describe a basic coupling technique, applicable to all kinds of
  processes, which allows trading memory use for a limited speedup. Next, I
  describe a specialized variant of it, which can be used to speed up the
  simulation certain kinds of processes satisfying a monotonicity criterion.
  This special algorithm increases the speed by several orders of magnitude with
  only a modest increase in memory usage.
\end{abstract}


\end{frontmatter}

\section{Introduction}\label{secIntro}

In its most basic form, coupling means constructing multiple stochastic
processes on the same underlying probability space \citep{liggett1999}.  In a
computer simulation, the role of the underlying probability space is played by
the random number generator, and the coupling techniques presented in this paper
can be described as simulating multiple realizations of a stochastic process in
parallel using the same stream of random events.

\label{secA}
The processes to which I will apply the simulation techniques described below
belong to the class of lattice contact processes introduced by
\citet{harris1974} as models of an endemic infection in a spatially structured
host population.  While the simulation techniques described in this
paper---particularly in \autoref{secB}---may in principle also be applied to
other interacting stochastic processes (like e.g.~the Ising process from
statistical physics), they were developed with the contact process in mind, and
it is the contact process which I will use to demonstrate them here.

In the basic lattice contact process, each site on a regular lattice represents
a single host individual, which, at any given time, may be in one of two states:
uninfected (0) or infected (1).  The state of the lattice, which consists of the
states of the individual sites, evolves as a continuous time Markov process.
\Figuref{algA} shows a conceptual illustration of the possible local transitions
which may occur: each infected site recovers independently with rate $r$ and
transmits the infection to a random neighbor site with rate $c$.\footnote{ For
  simplicity, I assume here that each site has the same number of neighbors;
  where that is not the case, it is common to take the transmission rate as
  proportional to the number of neighbors.}
\begin{figure}[tb]
  \centering
  \includegraphics[width=\columnwidth]{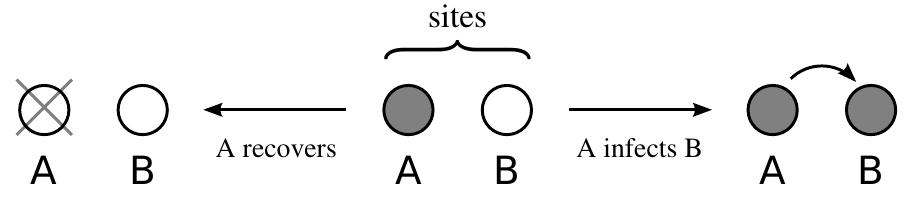}
  \caption{Possible local transition events in a basic contact process.  Empty
    circles denote uninfected and filled circles denote infected sites.}
  \label{figA}
\end{figure}

The word ``neighbor'' here should be understood in a broad sense: the algorithms
given in this paper work just as well regardless of which sites are considered
to be neighbors, and they can even be straightforwardly generalized to handle
arbitrary dispersal kernels or contact networks, where different pairs of sites
may have different probabilities of infecting each other.  Indeed, they can even
be adapted to simulate spatio-temporal point processes in continuous space.
Also, while I've followed the terminology of \citet{harris1974} in describing
the lattice sites as ``hosts'' which may be ``infected'' or ``uninfected'', they
can be more generally interpreted e.g.~as occupied or vacant habitat patches in
a stochastic patch occupancy model of metapopulation dynamics, or as individual
spatial loci in a lattice model of a plant population.  In \autoref{secE} I also
describe an application of these techniques to models with more than one
competing infector strain.

\begin{algorithm}[tbh]
  \caption{Naive contact process simulation.}
  \label{algA}
  \dontprintsemicolon
  \KwData{
    $S_x \in \{0,1\} \quad \forall x \in L = \{1, \dotsc, N\}$
  }
  \While{$t < t_{\max}$}{
    $A := \text{random site in }L$\;
    $x := \text{random number in }[0, r+c)$\;
    \eIf{$x < c$}{
      \textit{perform a contact event}: \\
      $B := \text{random neighbor of }A$\;
      $S_B \gets 1$ \textbf{if} $S_A = 1$\;
    }{
      \textit{perform a recovery event}: \\
      $S_A \gets 0$\;
    }
    \textit{advance clock by mean time between events}: \\
    $t \gets t + \frac{1}{N(r+c)}$\;
  }
\end{algorithm}
\Algoref{algA} shows a straightforward naive (i.e.~unoptimized) algorithm for
simulating the basic lattice contact process.  Here the array $S$ stores the
current state (0 for uninfected, 1 for infected) of each site on the lattice
$L$.  At every iteration, a single site $A$ on the lattice is randomly chosen as
the focus of an event, which will be a contact event if the uniform random
variable $x \in [0, r+c)$ is less than $c$ and a recovery event otherwise.
  After each iteration, the time $t$ is advanced by the expected time $1/(r+c)$
  between events per site, divided by the total number of sites $N$.\footnote{
    Strictly speaking, the time between consecutive events should, of course, be
    an exponentially distributed random variable, but replacing this random
    variable with its expectation is a commonly used simplification, justified
    by the fact that the expected difference between the approximate time so
    obtained and the ``true'' time scales as $\sqrt{t(N(r+c))^{-1}}$, and thus
    becomes relatively negligible compared to $t$ when $tN(r+c) \gg 1$
    (i.e.~after a large number of events).}

Experienced readers may note that there are several optimizations that could be
made to \autoref{algA}.  For example, it is wasteful to sample the focal site
$A$ from the entire lattice $L$, when we could instead maintain a list of the
currently infected sites and sample $A$ from that list.  For simplicity, I will
not include such well known optimizations in the example algorithms presented
here, nor will I discuss them except where relevant to the methods introduced
below.

\section{Coupling}\label{secB}

A general way to speed up \autoref{algA} is to simulate multiple coupled
instances of the process in parallel, as shown in \autoref{algB}.
\begin{algorithm}[tbh]
  \caption{General $n$-fold coupled contact process simulation.}
  \label{algB}
  \dontprintsemicolon
  \KwData{
    $S_{x,k} \in \{0,1\} \quad \forall x \in L = \{1, \dotsc, N\}, k \in K = \{1, \dotsc, n\}$
  }
  \BlankLine
  $c_{\max} := \max(c_1, \dotsc, c_n)$\;
  $r_{\max} := \max(r_1, \dotsc, r_n)$\;
  \While{$t < t_{\max}$}{
    $A := \text{random site in }L$\;
    $x := \text{random number in }[0, r_{\max}+c_{\max})$\;
    \eIf{$x < c_{\max}$}{
      \textit{perform a contact event}: \\
      $B := \text{random neighbor of }A$\;
      \For{$k \in \{1, \ldots, n\}$}{
        \lIf{$x < c_k$ {\rm\bf and} $S_{A,k} = 1$}{
          $S_{B,k} \gets S_{A,k}$\;
        }
      }
    }{
      \textit{perform a recovery event}: \\
      \For{$k \in \{1, \ldots, n\}$}{
        \lIf{$x-c_{\max} < r_k$}{
          $S_{A,k} \gets 0$\;
        }
      }
    }
    \textit{advance clock by mean time between events}: \\
    $t \gets t + \frac{1}{(r_{\max}+c_{\max})N}$\;
  }
\end{algorithm}

Here we have $n$ contact processes with contact and recovery rates $c_k$ and
$r_k$, $k \in K = \{1, \dotsc, n\}$ respectively, and instead of single states,
the array $S$ contains vectors of $n$ states $S_A = (S_{A,1}, S_{A,2}, \dotsc,
S_{A,n})$ for each site $A$.  The fixed $r$ and $c$ from \autoref{algA} are
replaced with $r_{\max} = \max_{k \in K} r_k$ and $c_{\max} = \max_{k \in K}
c_k$ respectively, but we update the $k$-th element of the state vector on each
contact event only if $x < c_k$, and on each recovery event only if $x-c_{\max}
< r_k$.  In this way, the effective contact and recovery rates for the $k$-th
process remain $c_k$ and $r_k$ respectively.

\begin{figure}[tb]
  \centering
  \includegraphics[width=\columnwidth]{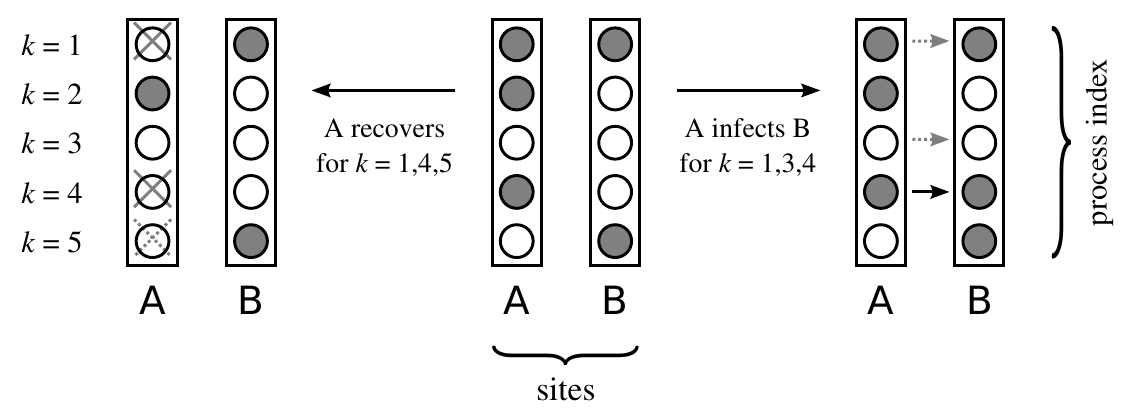}
  \caption{Examples of possible local transition events in a coupled contact
    process simulation.}
  \label{figB}
\end{figure}
\Figuref{algB} illustrates some example state transitions in \autoref{algB}.
Here the columns of circles represent the state vectors of two neighboring
sites.  Whenever a contact or recovery event is performed, the random variable
$x$ is used to determine the set of $k$ values for which the event in question
actually takes place, with the remaining elements in the state vectors being
unchanged.

Of course, the realizations of the processes simulated using \autoref{algB} will
not be independent, even though it is easy to see that each process has the
correct marginal transition probabilities when considered separately from the
others.  This lack of mutual independence must be kept in mind when interpreting
the results.
In particular, when sampling the behavior of a model over a range of parameter
values using traditional simulation methods like \autoref{algA}, a fairly common
practice is to use only one simulation run for each parameter value and to
instead increase the number of parameter values sampled, so that any stochastic
variation in the results will hopefully be evident as lack of correlation
between nearby sample points.  While using \autoref{algB} allows the process to
be simulated for many parameter values at once, the tradeoff is that any
stochastic variation in the results of a single run will be strongly correlated,
and so multiple runs will be necessary to properly estimate variances in the
results.
Even so, the number of runs of \autoref{algB} needed to obtain results of
comparable quality will often compare favorably with naively sampling the
parameter space using \autoref{algA}.

The performance gains for this basic coupling technique come mainly from the
reduction in loop overhead (i.e.~the time spent executing the looping
instructions, updating the clock variable, etc.) and random number generator
calls.  In particular, note that \autoref{algB} consumes just as many random
numbers to simulate $n$ coupled processes as \autoref{algA} needs to simulate
just one.  Random number generation can be a time-consuming process,
particularly if a high quality random number generator is used, and thus the
time savings from minimizing random number use can be substantial.
Depending on how the array $S$ is stored in memory, the locally sequential
memory access pattern of reading and writing one whole state vector at a time
may also be more efficient that the completely random pattern in \autoref{algA},
particularly if a compact representation of the state vectors is used.\footnote{
  In languages that don't have a native compactly stored bit vector datatype,
  vectors of 8, 16, 32 or 64 bits can be stored in suitably sized integer
  variables.  A convenient feature of such a representation is that the inner
  loops in \autoref{algB} can be replaced by bitwise Boolean logic operations
  such as $S_B \gets S_B \lor (S_A \land M(x))$, where the mask $M(x)$ can be
  efficiently looked up using e.g.~the square histogram method of
  \citet{marsaglia2004}.  For longer vectors, arrays or tuples of integers may
  be used.}

The down side, of course, is that no matter how compactly the state vectors are
represented, each vector of $n$ states needs at least $n$ bits of memory.  In
particular, the lattice size and the number of coupled processes needs to be
kept low enough that the entire lattice fits in the available memory without
swapping to disk, or else performance is likely to suffer catastrophically.  In
practice, some trial and error may be necessary to find the optimal number of
coupled processes to maximize overall simulation speed on a given system.

It is also worth noting that the speed of the coupled simulation depends on the
maximum rates of each type of event over the coupled processes.  If the
processes vary widely in their respective event rates, those with rates
significantly below the maximum will be simulated less efficiently than they
would be on their own, requiring the simulation to be run longer.  In such
cases, it may be more efficient to split the processes into groups with similar
event rates and to simulate each group separately.  Such grouping is
particularly recommended when using optimization techniques, such as those
mentioned at the end of \autoref{secA}, which depend on a substantial fraction
of the sites being entirely uninfected or entirely infected.

\section{Monotone coupling}\label{secC}

In some cases the general coupling technique described above can be made much
more efficient yet.  In particular, consider what would happen if we could order
the coupled processes in \autoref{algB} such that $c_i \le c_j$ and $r_i \ge
r_j$ for all $i < j$.  Then, if we also chose the initial condition so that, for
all $1 \le A \le N$ and $i < j$, $S_{A,i} \le S_{A,j}$, this condition would
still hold after one iteration of the main loop, and so would continue to hold
after any number of iterations.

Thus, we wouldn't need to keep track of the entire state vector $S_A$ for each
site $A$, but only of the lowest $k$ for which $S_{A,k} = 1$, allowing us to run
arbitrarily many coupled processes in parallel with much lower memory usage
(about $\log_2 n$ bits of memory per site) than with \autoref{algB} (which needs
$n$ bits per site).

More generally, let $S_A(p)$ denote the state of the site $A$ at a given time
under some family of coupled interacting stochastic processes parametrized by
the value $p$.  If the initial condition
\begin{equation}
  p \le p' \implies S_A(p) \le S_A(p') \quad \forall A
  \label{eqnMonotone}
\end{equation}
continues to hold under the time evolution of the process, we call the family of
processes monotone with respect to the parameter $p$.  If a family of processes
is monotone with respect to some parameter $p$, we can use the technique
described above to simulate them for \emph{all} values of $p$ (in a given range)
at the same time!

Obviously, not all interacting stochastic processes are monotone with respect to
a suitable parameter---indeed, monotonicity with respect to a parameter
necessarily implies that the process itself must be monotone with respect to its
initial condition \citep{liggett1999}, which many stochastic processes are not.
(For example, processes that exhibit cyclic dynamics are generally not
monotone.)  However, the basic contact process is indeed monotone, both in
itself and with respect to several parameters of interest, such as the contact
rate $c$ and the recovery rate $r$, as well as any parameter $p$ such that $c$
is an increasing and $r$ a decreasing function of $p$ (or \textit{vice versa}).

\begin{algorithm}[tbh]
  \caption{Monotone coupling simulation for the contact process with
    $0 \le r \le r_{\max}$, fixed $c$.}
  \label{algC}
  \dontprintsemicolon
  \KwData{
    $\theta_x \in [0,r_{\max}) \quad \forall x \in L = \{1, \dotsc, N\}$
  }
  \While{$t < t_{\max}$}{
    $A := \text{random site in }L$\;
    $x := \text{random number in }[0, r_{\max}+c)$\;
    \eIf{$x < c$}{
      \textit{A unconditionally infects B}: \\
      $B := \text{random neighbor of }A$\;
      $\theta_B \gets \max(\theta_A, \theta_B)$\;
    }{
      \textit{A recovers if $r \ge r_{\mathrm{crit}} = x-c$}: \\
      $\theta_A \gets \min(\theta_A, x-c)$\;
    }
    $t \gets t + \frac{1}{(r_{\max}+c)N}$\;
  }
\end{algorithm}
\begin{figure}[tb]
  \centering
  \includegraphics[width=\columnwidth]{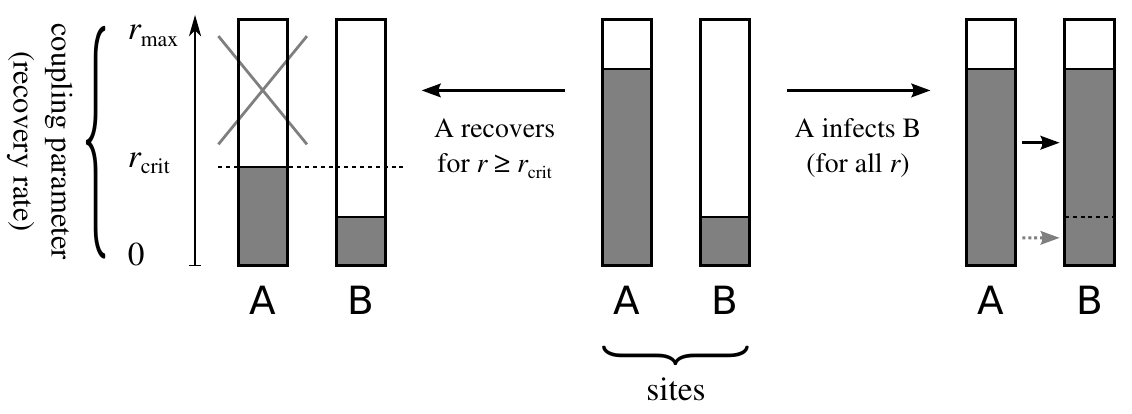}
  \caption{Examples of possible local transition events in a monotone coupled
    contact process simulation with $r$ as the coupling parameter.  The bars
    represent the state of a site, with the shaded area showing the parameter
    range for which that site is infected.  On each recovery event, a threshold
    value $r_{\mathrm{crit}}$ is chosen uniformly between $0$ and
    $r_{\max}$ and the focal site is marked as uninfected for all $r \ge
    r_{\mathrm{crit}}$.}
  \label{figC}
\end{figure}
As an example, \autoref{algC} shows how to simulate the basic contact
process for all values of $r \in [0, r_{\max})$ for a fixed $c$.  Here, the
state vector $S_A$ from \autoref{algB} is replaced by the threshold value
$\theta_A$, which stores the smallest value of the coupling parameter $r$ for
which the site $A$ is \emph{not} currently occupied.

\Figuref{algC} shows the types of events that occur in this simulation.  On
contact events, the threshold value $\theta_B$ of the target site $B$ is simply
raised up to the threshold $\theta_A$ of the infecting site $A$, showing that
$B$ is now infected in (at least) all the coupled processes in which $A$ was
infected before the event.  On recovery events, we reduce the threshold value
$\theta_A$ down to a random value $r_{\mathrm{crit}}$ uniformly chosen between $0$
and $r_{\max}$.  Conveniently, we already have such a random variable available:
the variable $x$ is uniformly distributed between $c$ and $r_{\max}+c$, so we
can simply let $r_{\mathrm{crit}} = x-c$.  In this way, only the fraction
$r/r_{\max}$ of all recovery events affect the state of the process with
parameter $r$.  Since the total rate of recovery events in \autoref{algC}
is $r_{\max}$ per site, the effective per site recovery rate for the process
with parameter $r$ is indeed $r$.

\label{secD}  
\begin{algorithm}[tbh]
  \caption{Monotone coupling simulation for the contact process with
    $0 \le c \le c_{\max}$, fixed $r$.}
  \label{algD}
  \dontprintsemicolon
  \KwData{
    $\theta_x \in [0,c_{\max}] \quad \forall x \in L = \{1, \dotsc, N\}$
  }
  \While{$t < t_{\max}$}{
    $A := \text{random site in }L$\;
    $x := \text{random number in }[0, r+c_{\max})$\;
    \eIf{$x < c_{\max}$}{
      \textit{A infects B if $c < c_{\mathrm{crit}} = x$}: \\
      $B := \text{random neighbor of }A$\;
      $\theta_B \gets \min(\theta_B, \max(x, \theta_A))$\;
    }{
      \textit{A recovers unconditionally}: \\
      $\theta_A \gets c_{\max}$\;
    }
    $t \gets t + \frac{1}{(r+c_{\max})N}$\;
  }
\end{algorithm}
\begin{figure}[tb]
  \centering
  \includegraphics[width=\columnwidth]{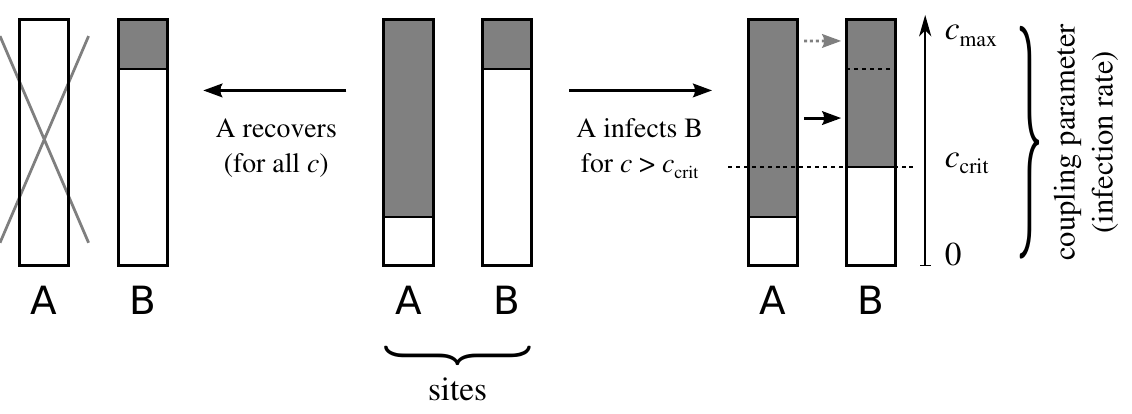}
  \caption{Examples of possible local transition events in a monotone coupled
    contact process simulation with $c$ as the coupling parameter.  The bars
    represent the state of a site, with the shaded area showing the parameter
    range for which that site is infected.  On each contact event, a threshold
    value $c_{\mathrm{crit}}$ is chosen uniformly between $0$ and
    $c_{\max}$ and the target site is marked as infected for all parameter
    values above $c_{\mathrm{crit}}$ for which the source site was infected before
    the event.}
  \label{figD}
\end{figure}
\Algoref{algD} simulates the same process for $c \in [0, c_{\max})$ and
fixed $r$.  Here $\theta_A$ denotes the smallest value of $c$ for which the site
$A$ is occupied; $\theta_A = c_{\max}$ means that the site $A$ is not currently
occupied in any of the coupled processes.
\Figuref{algD} illustrates the events possible in this algorithm: recovery
events occur independently of the coupling parameter $c$, while on contact
events, we choose a uniform random value $c_{\mathrm{crit}}$ between $0$ and
$c_{\max}$ (for which purpose, again, the uniform random variable $x$ is already
conveniently available) and lower the infection threshold $\theta_B$ of the
target site $B$ down to the maximum of $c_{\mathrm{crit}}$ and the infection
threshold $\theta_A$ of the focal site $A$.  Thus, a fraction
$c/c_{\mathrm{crit}}$ of all contact events affect a process with the coupling
parameter value $c$, giving that process the effective contact rate $c$.

\section{Multitype contact processes}\label{secE}

The monotone coupling technique is not restricted to the basic single-type
contact process.  The challenge in applying it to more complicated processes is
mainly in finding a suitable parameter which changes the dynamics in a
nontrivial, yet monotone, manner.  Fortunately, many processes do have such a
parameter, even if it may not always be the most interesting one.  If the goal
is to explore the entire parameter space, the existence of any monotone
parameter---even if trivial---allows much more efficient sampling of the
parameter space than if no such parameters exist.

For processes featuring two competing (and mutually exclusive) infectious
strains $a$ and $b$, with the respective infection and recovery rates $c_a$,
$c_b$, $r_a$ and $r_b$, a sufficient condition for the process to be monotone
with respect to a parameter $p$ is that $c_a$ and $r_b$ are (weakly) decreasing
and $r_a$ and $c_b$ (weakly) increasing functions of $p$.  If this condition
holds, we may order the possible states of a single lattice site as follows:
\begin{enumerate}
\item[1:] infected with strain $a$,
\item[2:] not infected,
\item[3:] infected with strain $b$,
\end{enumerate}
and arrange all transitions of the coupled process to preserve the monotonicity
property \eqref{eqnMonotone}.

Quite a few well known models possess such parameters.  For example, the
infection rates $\lambda_1$ and $\lambda_2$ in the original multitype contact
process defined by \citet{neuhauser1992} are both monotone parameters.  So is
the cost of altruism $C$ in the model of \citet{vanbaalen1998}.

In some cases, the same technique can be applied to processes with more than two
competing strains.  For example, in the three-strain model of
\citet{lanchier2006} for generalist--specialist coexistence on two site types,
the two specialist strains are each restricted to their respective site types,
and can thus be effectively treated as a single strain that cannot spread from
one site type to another.  In \citet{coex2env} I have applied the monotone
coupling technique to simulating a process essentially equivalent to that of
\citet{lanchier2006}, with the generalist infection probability $p \in [0,1]$ as
the coupling parameter.

\Algoref{algE} shows a simplified version of the algorithm used to simulate the
process studied in \citet{coex2env}.  (The original implementation also uses
occupancy / vacancy lists and various other optimizations.)
In this model, the sites are arbitrarily divided into two classes, and the
strains consists of two specialists, only capable of infecting sites of the
corresponding class, and a generalist strain capable of infecting either class
of sites with equal (but reduced compared to the specialists) probability.  No
superinfection is assumed to occur.  Thus, each site can be considered to be in
one of three states: infected by a specialist strain (1), uninfected (2), or
infected by the generalist strain (3).  (Each class of sites can only be
infected by one of the specialist strains, so it is not necessary to track which
specialist has infected a given site.)  With the site states numbered as above,
this process is monotone with respect to the infection probability $p \in [0,1]$
of the generalist strain.

\begin{algorithm}[tbh]
  \caption{Monotone coupled simulation for a multitype contact process
    with generalist infectivity $0 \le p \le 1$, fixed $c$ and $r$}
  \label{algE}
  \dontprintsemicolon
  \KwData{
    $\theta_x \in [0,1]^2 \quad \forall x \in L = \{1, \dotsc, N\}$
  }
  \While{$t < t_{\max}$}{
    $A := \text{random site in }L$\;
    $x := \text{random number in }[0, r+c)$\;
    \eIf{$x < c$}{
      $B := \text{random neighbor of }A$\;
      \If{$H_A = H_B$}{
        \textit{specialist always infects if host types match}: \\
        $\theta_{B,1} \gets \med( \theta_{B,1}, \theta_{A,1}, \theta_{B,2} )$\;
      }
      \textit{generalist infects if $p > x/c$}: \\
      $\theta_{B,2} \gets \med( \theta_{B,1}, \max(x/c, \theta_{A,2}), \theta_{B,2} )$\;
    }{
      \textit{$A$ recovers unconditionally}: \\
      $\theta_{A,1} \gets 0,$
      $\theta_{A,2} \gets 1$\;
    }
    $t \gets t + \frac{1}{N(r+c)}$\;
  }
\end{algorithm}
Here, $H_A \in \{0,1\}$ is the type of the site $A$ (which does not change
during the simulation), and $\theta_A = (\theta_{A,1}, \theta_{A,2})$ denotes
the threshold values of the coupling parameter $p$ at which the current state of
the site $A$ changes: for $p < \theta_{A,1}$ the site is infected by a
specialist, for $p \ge \theta_{A,2}$ it is infected by the generalist, and for
intermediate values of $p$ it is uninfected.

\begin{figure}[tb]
  \centering
  \includegraphics[width=\columnwidth]{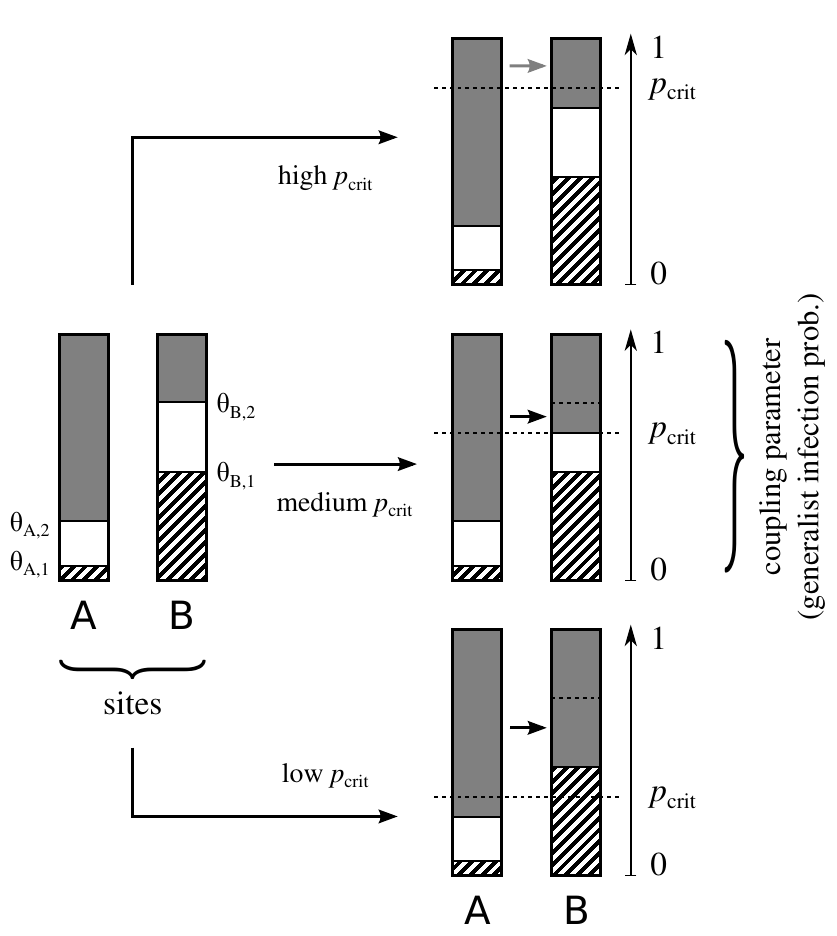}
  \caption{Some examples of possible infection events in a monotone coupled
    contact process simulation with two strains.  The striped and shaded regions
    denote the parameter ranges for which each site is infected by the strains
    $a$ and $b$.  The randomly chosen threshold $p_{\mathrm{crit}}$ gives the
    lowest value of the parameter $p$ for which the strain $b$ is transmitted.
    In the example, the focal site $A$ is infected by strain $b$ for all
    parameter values for which the target site $B$ is uninfected.  Thus,
    depending on whether $p_{\mathrm{crit}}$ lies above, within or below the
    uninfected range $[\theta_{B,1}, \theta_{B,2}]$, infection occurs for none,
    some or all parameter values within that range.}
  \label{figE}
\end{figure}
The function $\med(x, y, z)$ returns the median of its inputs, and is in effect
used to "clamp" $y$ to the interval $[x, z]$, so that, if $y$ lies outside the
interval, the closest endpoint is taken instead.  This is done so that, if the
effective infection threshold lies outside the uninfected range $[\theta_{B,1},
\theta_{B,2}]$ of the target site $B$, either no infection takes place or the
infection happens for the entire range, as shown in \autoref{figE}.  In
particular, this ensures that $\theta_{A,1} \le \theta_{A,2}$ stays true for all
sites $A$, provided that the initial state satisfies this.

\section{Efficient statistics collection}\label{secF}

The purpose of simulating a stochastic process is to collect some data about
it.
In some cases, the data we wish to collect, such as the presence or absence of a
certain strain, can be obtained simply from the final state of the process after
simulating it for $t_{\max}$ time units.  More commonly, however, we wish to
average some quantities, such as population densities, over a period of time, if
only to reduce the effects of stochastic fluctuations.

Of course, one way to accomplish this is to simply run the process for some time
interval $\delta_t$, sample the values of interest from the lattice state at
that point, and repeat until enough samples have been collected.  However, while
easy to implement, this method is somewhat inefficient.  If $\delta_t$ is small,
successive samples will be highly correlated, and so we will need to collect
many of them to average out temporal fluctuations.  Since each sampling step
usually involves iterating over the entire lattice, this can consume a lot of
time if done frequently.  On the other hand, if $\delta_t$ is large, we need to
simulate the process longer to collect a given number of samples, which also
consumes time.

Somewhere between these two extremes there presumably exists an optimal value of
$\delta_t$ (for a particular process) that minimizes the amount of computation
needed to achieve a given noise level.  However, rather than optimizing
$\delta_t$, what we'd really like to do would be to calculate the true average
of the values we're interested in over a given time interval while simulating
the process.

One way to do this is to observe that every infected site is sampled (at least)
twice during the course of each infection: once when the infection occurs, and
once when the site recovers.  If the mean infection length $\bar \tau = 1/r$ is
known and significantly shorter than the timespan $t$ which we are averaging
over, simply counting the number $n$ of successful infection (or recovery)
events during the interval and multiplying it with $\bar \tau / Nt$ (where $N$
is the number of sites in the lattice) will yield a very good estimate of the
average infection density over the chosen time interval.

Such counting can be easily incorporated into the naïve \autoref{algA} simply by
incrementing a (realization and strain specific) counter whenever the state of a
site changes (in the direction we are counting).  However, applying it to the
monotone coupled simulation algorithms in \autoref{secC} and \autoref{secE} is
slightly less trivial, since we are simulating many (conceptually infinitely
many) realizations of the process at the same time: obviously, storing a counter
for each of them and incrementing all those for which a change occurs would be
inefficient.

Instead, we can make use of the observation that, whenever a site state change
occurs in the coupled simulation algorithms, it is always over a contiguous
range of values of the coupling parameter (or possibly a small number of
disjoint ranges, as in \autoref{algE}).  Thus, to record the parameter range for
which an event occurs, it suffices to increment a counter corresponding to the
start of the range and to decrement the counter corresponding to its end.  We
can then obtain the number of events that have occurred for a given parameter
value simply by summing up the counters corresponding to parameter values below
it.

\begin{algorithm}[tbh]
  \caption{Variant of \autoref{algD} (monotone coupling simulation for the
    contact process with $0 < c < c_{\max}$, fixed $r$) with event counting.}
  \label{algF}
  \dontprintsemicolon
  \KwData{
    $\theta_x \in (0,c_{\max}] \quad \forall x \in L = \{1, \dotsc, N\}$\;
    $n_i \in \Z \quad \forall i \in \{1, \dotsc, \lceil c_{\max} / \delta_c \rceil\}$\;
    $m_i \in \Z \quad \forall i \in \{1, \dotsc, \lceil c_{\max} / \delta_c \rceil\}$\;
  }
  \While{$t < t_{\max}$}{
    $A := \text{random site in }L$\;
    $x := \text{random number in }[0, r+c_{\max})$\;
    \eIf{$x < c_{\max}$}{
      \textit{A infects B if $c < c_{\mathrm{crit}} = x$}: \\
      $B := \text{random neighbor of }A$\;
      $\theta_{\mathrm{new}} := \max(x, \theta_A)$\;
      \If{$\theta_B > \theta_{\mathrm{new}}$}{
        $n_{\lceil \theta_{\mathrm{new}} / \delta_c \rceil} \gets n_{\lceil \theta_{\mathrm{new}} / \delta_c \rceil} + 1$\;
        $n_{\lceil \theta_B / \delta_c \rceil} \gets n_{\lceil \theta_B / \delta_c \rceil} - 1$\;
        $\theta_B \gets \theta_{\mathrm{new}}$\;
      }
    }{
      \textit{A recovers unconditionally}: \\
      $m_{\lceil \theta_A / \delta_c \rceil} \gets m_{\lceil \theta_A / \delta_c \rceil} + 1$ {\rm\bf if} $\theta_A < c_{\max}$\;
      $\theta_A \gets c_{\max}$\;
    }
    $t \gets t + \frac{1}{(r+c_{\max})N}$\;
  }
\end{algorithm}
\Algoref{algF} shows a variant of \autoref{algD} with this event counting
implemented for both infection and recovery events.  The parameter range $(0,
c_{\max})$ is divided into segments of length $\delta_c$, with the array $n_1,
\dotsc, n_{\lceil c_{\max} / \delta_c \rceil}$ of counters counting infection
events while the array $m_1, \dotsc, m_{\lceil c_{\max} / \delta_c \rceil}$
counts recovery events.  (Of course, in practice one would only track one of
these event types, since the difference between the event counts can be more
easily obtained simply by counting the number of infected sites before and after
the time interval of interest.)  At the end of the simulation, $\sum_{i=1}^k
n_i$ gives the number of infection events, and $\sum_{i=1}^k m_i$ the
corresponding number of recovery events, that have occurred for the parameter
value $c = k \delta_c$.

Of interest is the fact that, as recovery events in \autoref{algD} are always
unconditional, we need not record the endpoint of the affected range, since it
is always $c_{\max}$.  This saves time, and makes tracking recovery events more
convenient for this particular coupling parameter.  The same holds for the
coupled multitype process simulated in \autoref{algE}, where recovery events are
also unconditional.  There, tracking infection events would require four counter
adjustments per event, while tracking recovery events requires only two.

\section{Discussion}\label{secDiscuss}

Even though the basic principle of simulating interacting stochastic processes
on a computer is simple and straightforward, doing it efficiently can be
surprisingly complicated.  The coupled simulation technique described here, and
in particular the monotone coupling technique described in \autoref{secC} and
\autoref{secE}, are useful tools that can substantially speed up the simulation
of certain common types of contact processes by allowing the simulation of many
parametrized variants of the process at the same time.

There is no single algorithm that would be optimal for simulating all possible
interacting stochastic processes (or, at least, the author is not aware of any
such thing).  The usefulness and applicability of the techniques described in
this paper will have to be individually evaluated for each particular class of
processes one is interested in simulating.

This paper does not attempt to provide a comprehensive description of all the
optimization techniques available for simulating interacting stochastic
processes.  There are several well known optimization techniques, such as the
occupancy lists briefly mentioned in \autoref{secA}, which are more or less
orthogonal to the techniques presented in this paper and can (and should) be
combined with them where applicable.

While I have mainly used the classical lattice contact process of
\citet{harris1974} as the canonical example with which to demonstrate these
simulation techniques, \autoref{secE} demonstrates that, even though the
monotone coupling technique in particular requires certain rather specific
properties of the process, they are nonetheless applicable to a wider range of
ecological models.

Although this paper is written mainly with discrete lattice models in mind,
there seems to be no reason why the techniques described here could not be
naturally adapted to simulations of spatio-temporal point processes in
continuous space \citep{dieckmann1997}.  Indeed, such processes can be viewed as
a limiting case of lattice models as the number of sites per area tends to
infinity.  Of course, with infinitely many sites, sampling from the entire set
of sites will not be feasible, so that the use of an occupancy list, which was
mentioned as an optimization technique in \autoref{secA}, becomes a necessary
part of the simulation algorithm.  Also, to implement density regulation, the
handling of contact events must be modified to account for the suppression of
offspring growth in areas close to existing individuals.  Still, these are both
standard features of any spatio-temporal point process simulation algorithm, and
should not interfere with the coupling technique in any way.

\pagebreak[3]
\bibliographystyle{model2-names}
\bibliography{../../references}

\end{document}